\documentclass[journal=nalefd,manuscript=letter,amsmath,amssymb,superscriptaddress]{achemso}
\setkeys{acs}{articletitle = true,etalmode=truncate, maxauthors=10}
\usepackage{graphicx}
\usepackage{amssymb}
\usepackage{bm}
\usepackage{tabularx}

\usepackage{amsmath}
\usepackage[dvipsnames]{xcolor}

\usepackage[colorlinks=true,
  		linkcolor=black,
	        citecolor=blue,
	        urlcolor=blue]{hyperref}

\newcommand{\ie}{\emph{\ie, }}

\newcommand{\dqmp}{Department of Quantum Matter Physics, University of Geneva, 24 Quai Ernest Ansermet, CH-1211 Geneva, Switzerland}
\newcommand{\gap}{Group of Applied Physics, University of Geneva, 24 Quai Ernest Ansermet, CH-1211 Geneva, Switzerland}	
\newcommand{\IC}{Universit\'e Paris Est Creteil, CNRS, ICMPE, UMR 7182, F-94320, Thiais, France}

\keywords{van der Waals heterostructures, semimetal, band overlap, ionic liquid gating, 2D materials\\}
\title{Synthetic Semimetals with van der Waals Interfaces}
\date{\today}
\author{Bojja Aditya Reddy}
\altaffiliation{These two authors contributed equally}
\affiliation{\dqmp}
\alsoaffiliation{\gap}

\author{Evgeniy Ponomarev}
\altaffiliation{These two authors contributed equally}
\affiliation{\dqmp}
\alsoaffiliation{\gap}

\author{Ignacio Guti\'errez-Lezama}
\affiliation{\dqmp}
\alsoaffiliation{\gap}
\author{Nicolas Ubrig}
\affiliation{\dqmp}
\alsoaffiliation{\gap}
\author{C\'eline Barreteau}
\affiliation{\dqmp}
\alsoaffiliation{\IC}
\author{Enrico Giannini}
\affiliation{\dqmp}
\author{Alberto F. Morpurgo}
\affiliation{\dqmp}
\alsoaffiliation{\gap}
\email{alberto.morpurgo@unige.ch}

\AtBeginDocument{\usepackage{booktabs}}
\begin{document}

\begin{abstract}
The assembly of suitably designed van der Waals (vdW) heterostructures represents a new approach to produce artificial systems with engineered electronic properties. Here, we apply this strategy to realize synthetic semimetals based on vdW interfaces formed by two different semiconductors. Guided by existing ab-initio calculations, we select WSe$_2$ and SnSe$_2$ mono and multilayers to assemble vdW interfaces, and demonstrate the occurrence of  semimetallicity by means of different transport experiments. Semimetallicity manifests itself in a finite minimum conductance upon sweeping the gate over a large range in ionic liquid gated devices, which also offer spectroscopic capabilities enabling the quantitative determination of the band overlap. The semimetallic state is additionally revealed in Hall effect measurements by the coexistence of electrons and holes, observed by either looking at the evolution of the Hall slope with sweeping the gate voltage or with lowering temperature. Finally, semimetallicity results in the low-temperature  metallic conductivity of interfaces of two materials that are themselves insulating. These results demonstrate the possibility to implement a state of matter that had not yet been realized in vdW interfaces, and represent a first step towards using these interfaces to engineer topological or excitonic insulating states.
\end{abstract}


The possibility to assemble heterostructures of different two-dimensional (2D) \emph{van der Waals} (vdW) materials\cite{geim_van_2013} without any constraint imposed by the need to match their lattices  discloses unprecedented opportunities to realize artificial systems\cite{novoselov_2d_2016,iannaccone_quantum_2018} hosting novel electronic states. Known examples include the modification of the band structure of graphene placed onto hexagonal boron nitride\cite{yankowitz_emergence_2012,dean_hofstadters_2013,ponomarenko_cloning_2013}, the drastic enhancement of spin-orbit interaction in graphene on semiconducting transition metal dichalcogenide (TMD) substrates \cite{wang_strong_2015,wang_origin_2016,benitez_strongly_2018,island_spinorbit-driven_2019,wang_quantum_2019,Avsar2019}, the occurrence of superconductivity in twisted bilayer graphene \cite{cao_unconventional_2018,lu_superconductors_2019,yankowitz_tuning_2019,codecido_correlated_2019}, and the appearance of correlated insulating states in gate-doped twisted TMD bilayers\cite{wu_topological_2019,Wang2019}. Many more interesting artificial systems can be envisioned, because vdW interfaces allow 2D materials hosting distinct electronic phenomena --such as superconductivity\cite{xi_ising_2016,sharma_2d_2018}, magnetism \cite{gong_discovery_2017,huang_layer-dependent_2017, deng_gate-tunable_2018,song_giant_2018,wang_very_2018,gibertini_magnetic_2019}, charge density waves \cite{terashima_charge-density_2003,soumyanarayanan_quantum_2013,ryu_persistent_2018}, and more -- to be brought in contact, within sub-nanometer distances.

Here we demonstrate how vdW interfaces can be used to realize synthetic semimetals, by stacking on top of each other two atomically thin semiconductors with a suitable band alignment. The strategy is similar to that employed recently to assemble vdW interfaces acting as artificial semiconductors\cite{lee_atomically_2014,ross_interlayer_2017,ponomarev_semiconducting_2018} with a controlled band-gap (see Fig. 1(a)). Realizing a semimetallic state requires the bottom of the conduction band in one of the semiconductors to be lower in energy than the top of the valence band in the other (see shaded bands profiles in Fig. 1(b)), so that the conduction and valence bands of the two materials overlap in energy. Under these conditions, charge is transferred from one material to the other, causing both bands to be partially filled. The Fermi level is then positioned in the energy interval in which the two bands overlap, turning the interface of two insulating 2D materials into a conductor, due to the simultaneous presence of  electrons and holes.

Semimetals of this type are expected to exhibit  interesting phenomena, depending on details of the interfacial electronic structure. If the overlapping bands of the constituent materials are both centered at the $\Gamma$-point, for instance, the interface should host a 2D quantum spin Hall state. Indeed, in this case the situation is analogous to that of III-V InAs/GaSb heterostructures\cite{knez_evidence_2011,suzuki_edge_2013,du_robust_2015}, with the advantage that in vdW interfaces, the atomic thickness of the constituent 2D materials can strongly enhance wavefunction overlap and hybridization effects between the two layers, resulting in a much larger gap between the inverted bands (for this to occur, strong spin-orbit interaction needs to be present, which is the case for the 2D materials that we consider\cite{zhu_giant_2011,xiao_coupled_2012,kosmider_large_2013}). If the band overlap does not occur at the same position in $k$-space, coupling between electrons and holes in the two layers can still occur, but is predominantly mediated by Coulomb interaction. In this regime, theory\cite{jerome_excitonic_1967} predicts the system to undergo a transition to an excitonic insulating state, in which electrons and holes are bound together, forming neutral excitons that do not carry current. Such an excitonic insulating state has not been observed to occur spontaneously in natural semiconducting materials, and vdW interfaces offer a new platform for its search.

\begin{figure}
    \centering
    \includegraphics[width= 0.7 \textwidth]{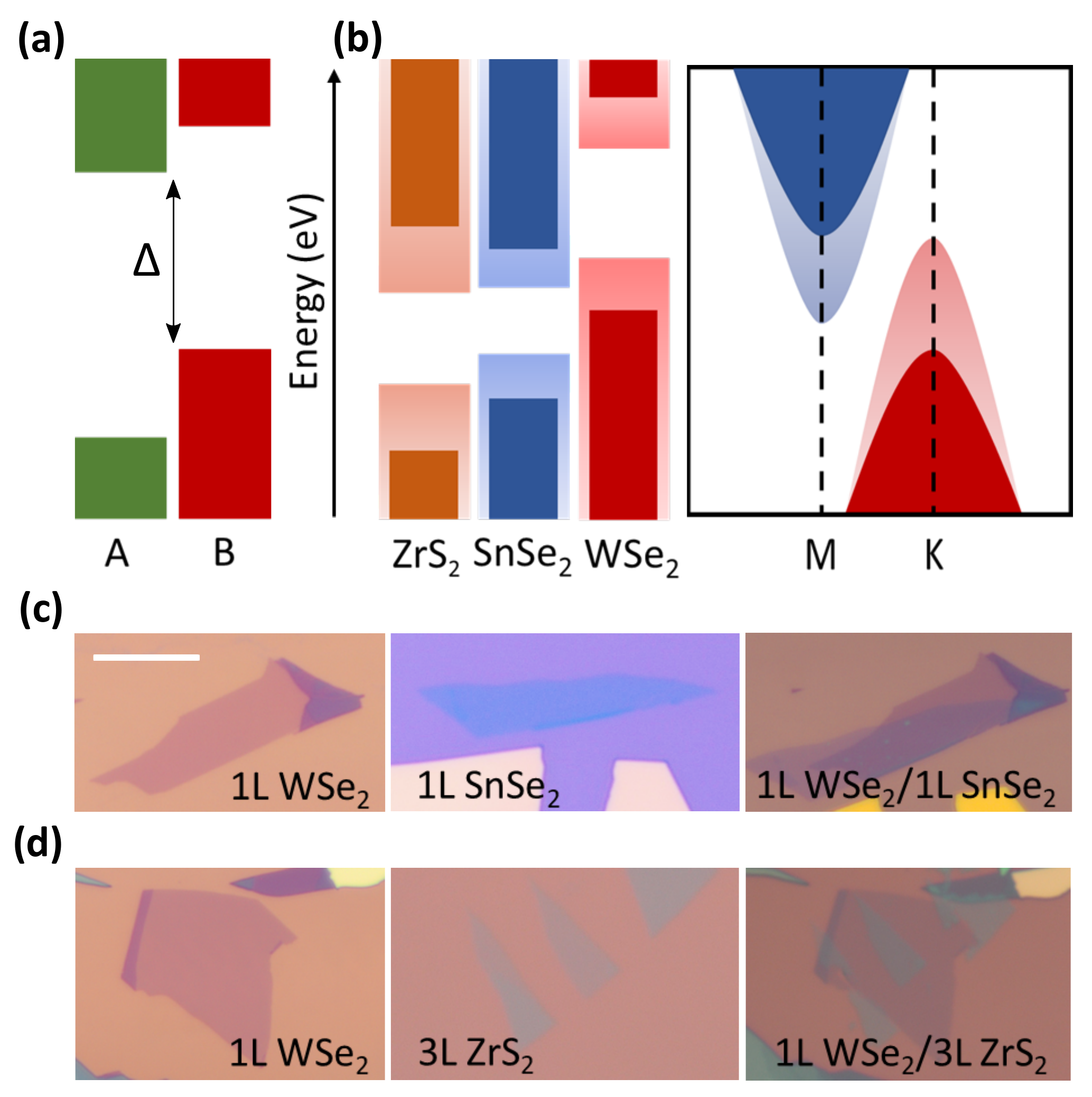}
  \caption{(a) Schematics of the bands of an artificial semiconductor, i.e., a vdW interface formed by 2D semiconductors A and B, such that at the interface the conduction band is determined by A and the valence band by B. (b) Left: expected band alignment at interfaces formed by WSe$_2$ and either ZrS$_2$ or SnSe$_2$. According to Refs \citenum{zhang_systematic_2016} and \citenum{koda_trends_2018} these systems exhibit either a band overlap (shaded bands in the diagram), or a small gap (full color bands in the diagram). Right: schematics of band dispersion for SnSe$_2$, ZrS$_2$ and WSe$_2$, showing that the maximum and minimum of the valence and conduction band occur at different position in $k$-space (shaded and full bands --exhibiting either a band overlap or a small gap--  correspond to the prediction of Refs \citenum{zhang_systematic_2016} and  \citenum{koda_trends_2018}. The inset shows the schematics of the Brillouin Zone of these materials. (c) and (d) Optical microscope images of individual semiconductor and their of interfaces, as indicated in the legends of each image (the scale bar is 10 $\mu$m).}
  \label{fig:Fig1}
\end{figure}

Our goal here is to show that a semimetallic state can indeed be realized in vdW interfaces formed by two different 2D semiconductors, and to demonstrate experimentally all its basic characteristic features: the presence of band overlap, the coexistence of electrons and holes, and the metallic nature of transport. Identifying suitable constituent semiconductors to assemble a semimetallic vdW interface is not straightforward, because for most semiconducting 2D materials the relative band offset is too small to bridge their band gap and create a band overlap. To guide our search, we rely on \textit{ab-initio} calculations reported in the literature, discussing the electronic structure of interfaces based on  combinations of monolayers of many different vdW semiconductors. In particular, the results of Zhang et al.\cite{zhang_systematic_2016} and Koda et al.\cite{koda_trends_2018} suggest that monolayers of WSe$_2$ and SnSe$_2$ or of WSe$_2$ and ZrS$_2$ should yield either a band overlap or a small band gap (see Fig. 1(b)). If sufficiently small, the presence of a gap between two monolayers does not necessarily represent a problem, because selecting bi or trilayers of these same materials can then likely allow a finite band overlap at the interface to be achieved (simply because the band gap of most 2D semiconductors under consideration decreases significantly upon increasing their thickness). These combinations of 2D materials are therefore promising to realize synthetic semimetallic systems. Nevertheless, since the precision of ab-initio calculations for the quantitative determination of material band gaps and of their band alignment is limited, it remains to be seen whether experimentally these interfaces behave as predicted theoretically (indeed, in recent work where a band overlap was expected based on theoretical predictions, no sign of a semimetallic state was found in the experiments\cite{kashiwabara_electrical_2019}).

To verify experimentally the occurrence of an interfacial semimetallic state we perform different types of transport measurements. To this end, we realize devices with nano-fabricated metallic contacts attached to the interfaces, and equip them with ionic liquid gates, to allow the accumulation of very large densities of charge carriers (either electrons or holes). Thanks to their very large geometrical capacitance these devices also enable spectroscopic measurements, as we explain in more details below. Examples of exfoliated layers of WSe$_2$, SnSe$_2$ and ZrS$_2$ used in our experiments are shown in Fig. 1(c) and (d), together with the images of the resulting vdW interfaces. Fig. 2 shows both the schematics of ionic liquid gated field-effect transistor (FET) devices (Fig. 2(c)) as well as optical microscope images of different interface devices with attached metallic contacts (insets of Fig. 2(d) and (e)). In practice, devices were realized using exfoliated layers of different thickness, ranging from monolayer (1L) to six-layers (6L) (the sources of the bulk crystals used for exfoliation are listed in the supplementary information). Exfoliation and interface assembly were done with the aid of a motorized stage in a nitrogen-filled glove box ($<$ 0.1 ppm O$_2$ and H$_2$O) in order to avoid any material deterioration (for ZrS$_2$ this is essential, as processing this material in air causes substantial degradation). The procedures used to assemble the interfaces, to attach contacts, and to apply the ionic liquid (P14-FAP) are by now rather conventional and are described in detail in the supplementary information.

\begin{figure*}
    \includegraphics[width= 0.8\textwidth]{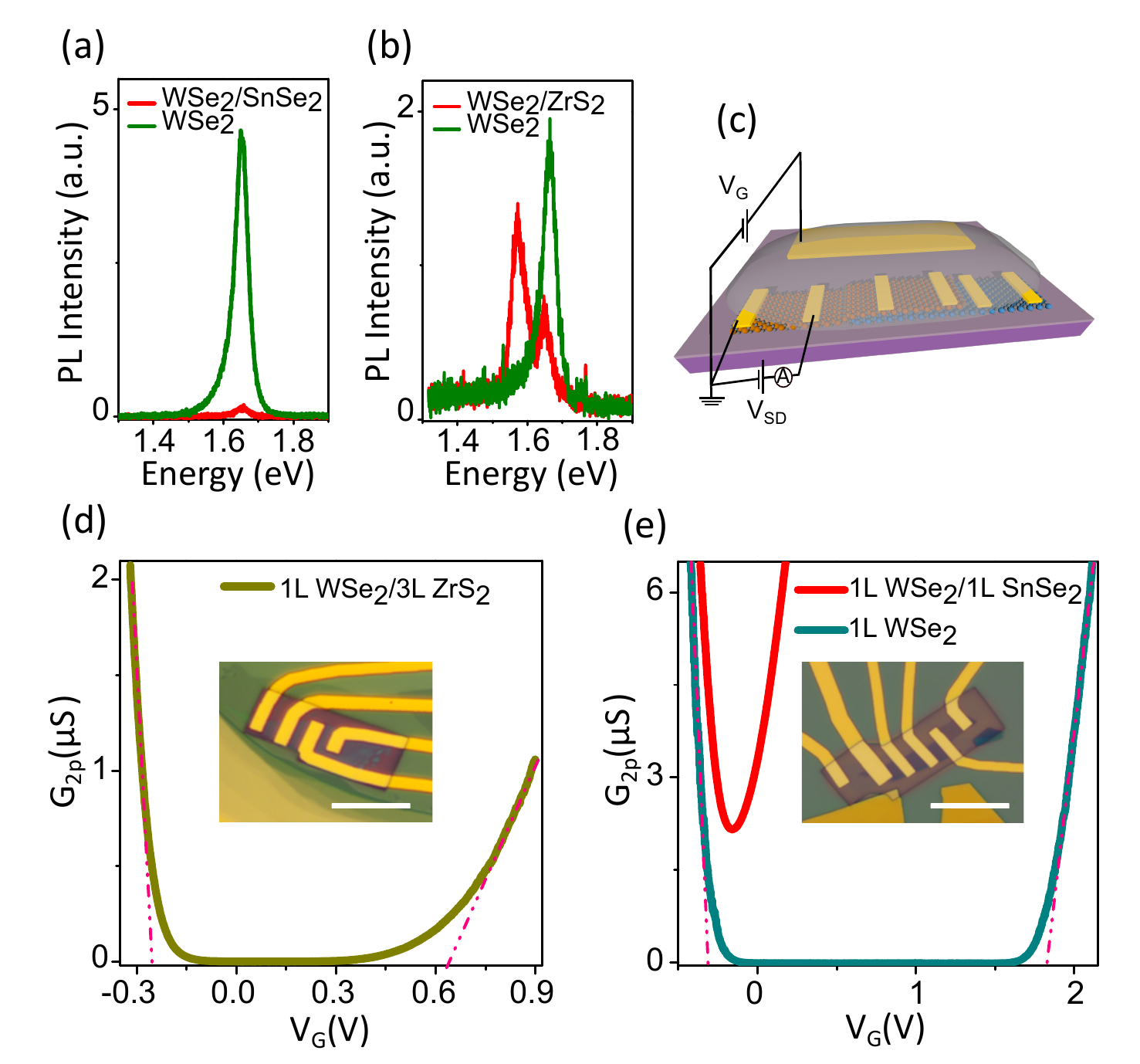}
  \caption{ (a) Photoluminescence (PL) spectra of a 1L-WSe$_2$/1L-SnSe$_2$ interface (red curve) and of the individual constituent 1L-WSe$_2$ layer (green curve). In both spectra, the peak at 1.65 eV originates from recombination of the A-exciton in 1L-WSe$_2$. As compared to the isolated constituent 1L-WSe$_2$, in the interface the PL intensity is nearly completely quenched. (b) PL spectra of a 1L-WSe$_2$/3L-ZrS$_2$ (red curve) and of the isolated constituent 1L-WSe$_2$ (green curve). In this case, the intensity of the PL from the interface and from the constituent monolayer are comparable (the double peak observed in the interface region likely originates from trion formation). (c) Schematics of an ionic-liquid gated device, with source and drain contacts on all the separate parts of the structure (i.e., the two constituent semiconductors and their interface). The schematics of the electrical circuit used to perform FET measurements is also shown. (d) Two-terminal conductance ($G_{2p}$) as a function of ionic-liquid gate voltage ($V_G$) in a 1L-WSe$_2$/3L-ZrS$_2$ interface, exhibiting ambipolar transport. (e) Two-terminal conductance ($G_{2p}$) as a function of ionic-liquid gate voltage ($V_G$) measured on a 1L-WSe$_2$/1L-SnSe$_2$ interface (red curve) and on the constituent 1L-WSe$_2$ (cyan curve). In (d) and (e) the red dash-dotted lines indicate how the respective transfer curves are extrapolated to extract the threshold voltages for electron and hole conduction. The insets in (d) and (e) show optical microscope images of the devices used in the measurements (the scale bar is 10 $\mu$m). All measurements were done at room temperature (since at below 270K the ionic liquid freezes).}
  \label{fig:Fig2}
\end{figure*}

As a first characterization step, for some of  the as-assembled interfaces we performed photoluminescence (PL) spectroscopy measurements  prior to attaching electrical contacts. PL spectra measured on 1L-WSe$_2$/1L-SnSe$_2$ and 1L-WSe$_2$/3L-ZrS$_2$ are shown in Fig. 2(a) and 2(b) (red lines), together with the spectrum of the WSe$_2$ layer in the corresponding samples (green lines; we have not detected any measurable PL signal from neither the SnSe$_2$ nor the ZrS$_2$ layers, both of which are indirect gap semiconductors). For the 1L-WSe$_2$/1L-SnSe$_2$ sample a nearly complete quenching of the PL signal is seen in the interface region (compare the red and the green line in Fig. 2(a)), indicating the presence of a non radiative decay path for photo-excited electron-hole pairs. This should indeed be expected if the band alignment gives rise to a semimetallic interfacial state, since transferring of carriers from the valence band of WSe$_2$ to the conduction band of SnSe$_2$  prevents radiative recombination within the WSe$_2$ layer. In the WSe$_2$/ZrS$_2$ interface, instead, the intensity of the WSe$_2$ peak has comparable spectral weight as in the bare WSe$_2$ layer (Fig. 2(b); in the interface the peak is split, likely because the presence of some excess charge allows the formation of trions). Therefore, PL measurements provide a first indication  that the band alignment in WSe$_2$/SnSe$_2$ --but not in WSe$_2$/ZrS$_2$-- may allow the realization of an interfacial semimetallic state.

Important indications as to the nature of the interfacial band structure can be obtained by transport measurements as a function of gate voltage, in transistor devices gated with an ionic liquid (see Fig. 2(c)). In such a device, a change in gate voltage ($\delta V_G$) induces both a change in chemical potential ($\delta E_F$) and in electrostatic potential  ($\delta \phi$), which in the geometry of a parallel plate capacitor is given by $\delta \phi = \frac{e^2 \delta n}{C_G}$ ($\delta n$ is the induced variation of accumulated charge density and $C_G$ the geometrical capacitance between gate and FET channel). These different quantities are related as:
\begin{equation}
    e\delta V_G = \delta E_F +e\delta \phi = \delta E_F + \frac{e^2 \delta n}{C_G}
\end{equation}
For an ionic liquid FET, the gate capacitance $C_G$ is extremely large. If the Fermi energy is swept through the material band gap, so that $\delta n \simeq 0$ (in the absence of defects, $\delta n$ nominally vanishes, since there are no states available inside the gap to accumulate charge),  the last term $\frac{e^2 \delta n}{C_G}$ can be neglected. It follows that in this regime $e\delta V_G = \delta E_F$, i.e., a change in gate voltage directly corresponds to a change in position of the Fermi level. It is this relation that enables the use of ionic liquid gated FETs to perform spectroscopic measurements. Specifically, when $V_G$ equals the threshold voltages $V_{th}^e$/$V_{th}^h$ for electron/hole accumulation, the Fermi level is located at the conduction/valence band edge, and we have $e(V_{th}^e-V_{th}^h)=\Delta$. That is: the difference in threshold voltages for electron and hole accumulation provides a direct measurement of  the material band gap. We have repeatedly used this technique in the past to measure the band gap of mono and multilayers of many different semiconducting TMDs\cite{braga_quantitative_2012,lezama_surface_2014,jo_mono-_2014,gutierrez-lezama_electroluminescence_2016} (as well as their band offsets\cite{ponomarev_semiconducting_2018}) and obtained precise and reproducible result. Here, we apply the same technique to WSe$_2$/SnSe$_2$ and WSe$_2$/ZrS$_2$ and their constituents (see Fig. 2(d,e) and Fig. (3))

To illustrate the principle of the technique, we start by discussing measurements on ionic liquid gated WSe$_2$ monolayer FET, nominally identical to devices that have been used in the past to determine the band gap of this material\cite{ponomarev_semiconducting_2018,zhang_band_2019}. The two-terminal conductance $G_{2p}$ measured as a function of $V_G$ (see cyan curve in Fig. 2(e)) shows clear ambipolar transport. The conductivity is high for negative $V_G$, when the Fermi level is located in the valence band of 1L-WSe$_2$, it vanishes in an extended interval of gate voltage (corresponding to having $E_F$ located in the band gap of 1L-WSe$_2$), and increases again as $V_G$ is sufficiently large to shift $E_F$ into the conduction band. The dashed-dotted lines show how the linear parts of the $G_{2p}(V_G)$ are extrapolated to zero, to extract the threshold voltages for electron and hole conduction. From their difference we obtain the band gap of 1L-WSe$_2$, $\Delta_{WSe2}\approx 2.1$ eV (if the analysis is performed by plotting $G_{2p}$ as a function of reference potential and not of $V_G$ --as we discussed multiple times elsewhere\cite{braga_quantitative_2012,ponomarev_ambipolar_2015,ponomarev_hole_2018}-- the quantitatively precise value $\Delta_{WSe2}\simeq 1.9$ eV\cite{zhang_band_2019} is obtained; see supplementary information). The exact same procedure can be applied to the interfacial regions, enabling the interface band gap to be determined\cite{ponomarev_semiconducting_2018}. Fig. 2(d) shows analogous $G_{2p}(V_G)$ data measured on a 1L-WSe$_2$/3L-ZrS$_2$, showing a behavior qualitatively similar to that of 1L-WSe$_2$, i.e.,  $G_{2p}(V_G)$ vanishes over an extended interval of gate voltages. This implies the existence of a finite band gap at the interface $\Delta_{WSe2-ZrS2}\approx 0.9$ eV and makes us conclude that WSe$_2$/ZrS$_2$ does not exhibit semimetallicity.

\begin{figure}
    \centering
    \includegraphics[width= 0.6 \textwidth]{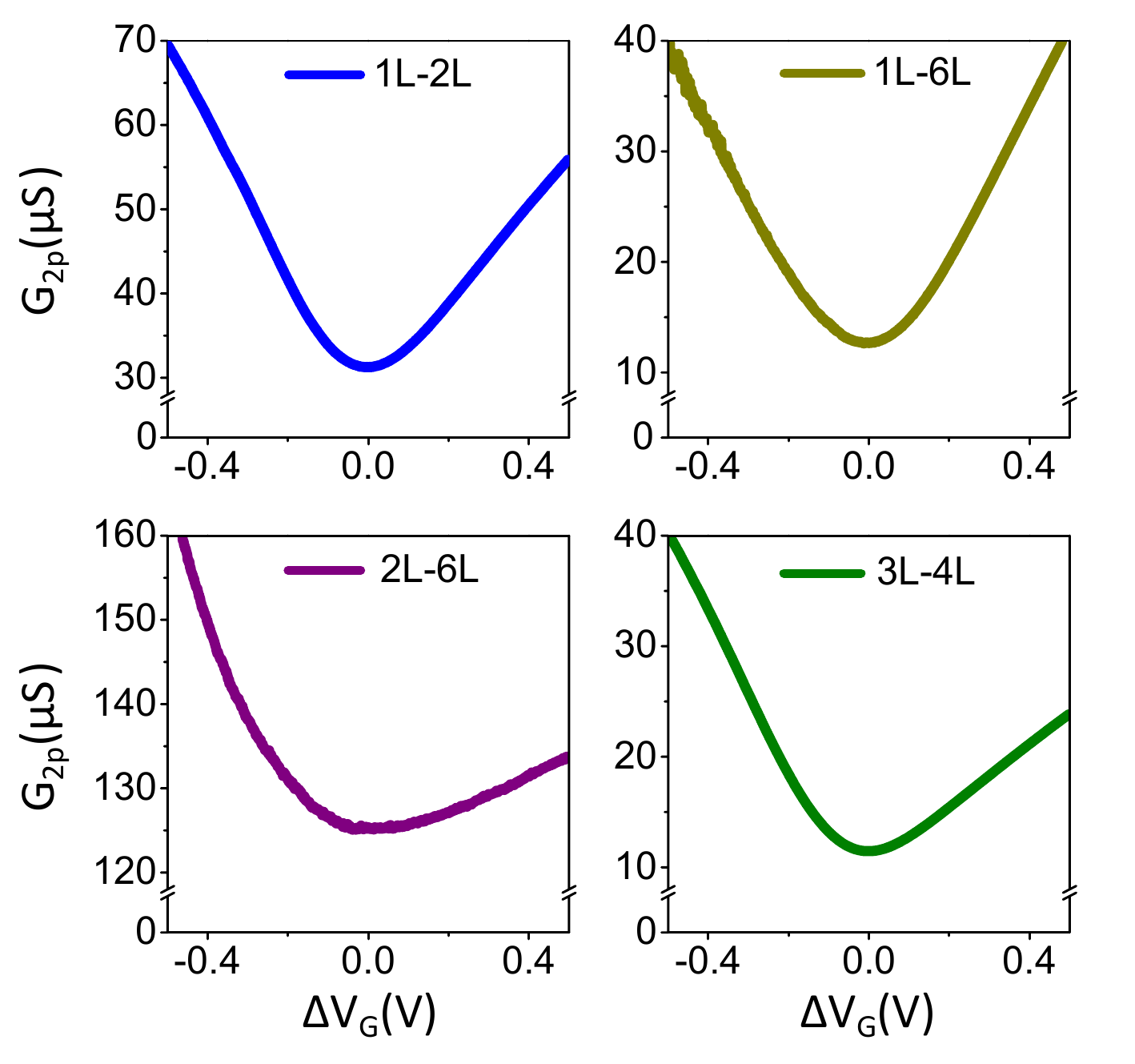}
  \caption{ Two-terminal conductance $G_{2p}$ as a function of $\Delta V_G$ = $V_G - V_{Gmin}$, the difference between applied gate voltage ($V_G$) and the gate voltage corresponding to the conductance minimum ($V_{Gmin}$), for WSe$_2$/SnSe$_2$ interfaces with different layer thickness (as indicated in the legend). The data show that the observed behavior is very robust (as measurements are done in a two-terminal configuration, a systematic analysis of the value of the conductance at the minimum is prevented by the possible presence of a contact resistance). All measurements were done at room temperature, to avoid freezing of the ionic liquid.}
  \label{fig:Fig3}
\end{figure}

The behaviour observed in 1L-WSe$_2$/1L-SnSe$_2$ interfaces is distinctly different, as illustrated by their two probe conductivity $\sigma_{2p}$ that remains finite for all values of $V_G$. Fig. 2(e) shows that for negative $V_G$ the conductivity nearly overlaps with the conductivity of the WSe$_2$ monolayer  measured on the same device (the structure contains multiple contacts that enable measuring separately the WSe$_2$/SnSe$_2$ interface and the WSe$_2$ layer, see the inset of Fig. 2(e)), it reaches a minimum value of $G_{2p}\simeq 2 \mu$S, and then increases again for positive $V_G$ values. No extended interval of gate voltage is present in which $G_{2p}$ vanishes as observed in WSe$_2$ (see cyan curve in Fig. 2(e)) or in  WSe$_2$/ZrS$_2$ interfaces (see Fig. 2(d)). The observed qualitative behavior --expected from a semimetal or a zero-gap semiconductor such as graphene-- is robust and was observed in all WSe$_2$/SnSe$_2$ interfaces that we studied, irrespective of the thickness of the constituent layers, as illustrated by the measurements performed on different devices shown in Fig. 3. These observations provide a direct indication of the absence of a sizable band gap in WSe$_2$/SnSe$_2$, confirming the conclusion suggested by the quenching of PL observed in 1L-WSe$_2$/1L-SnSe$_2$ discussed above.

\begin{figure}
    \centering
    \includegraphics[width= 0.6 \textwidth]{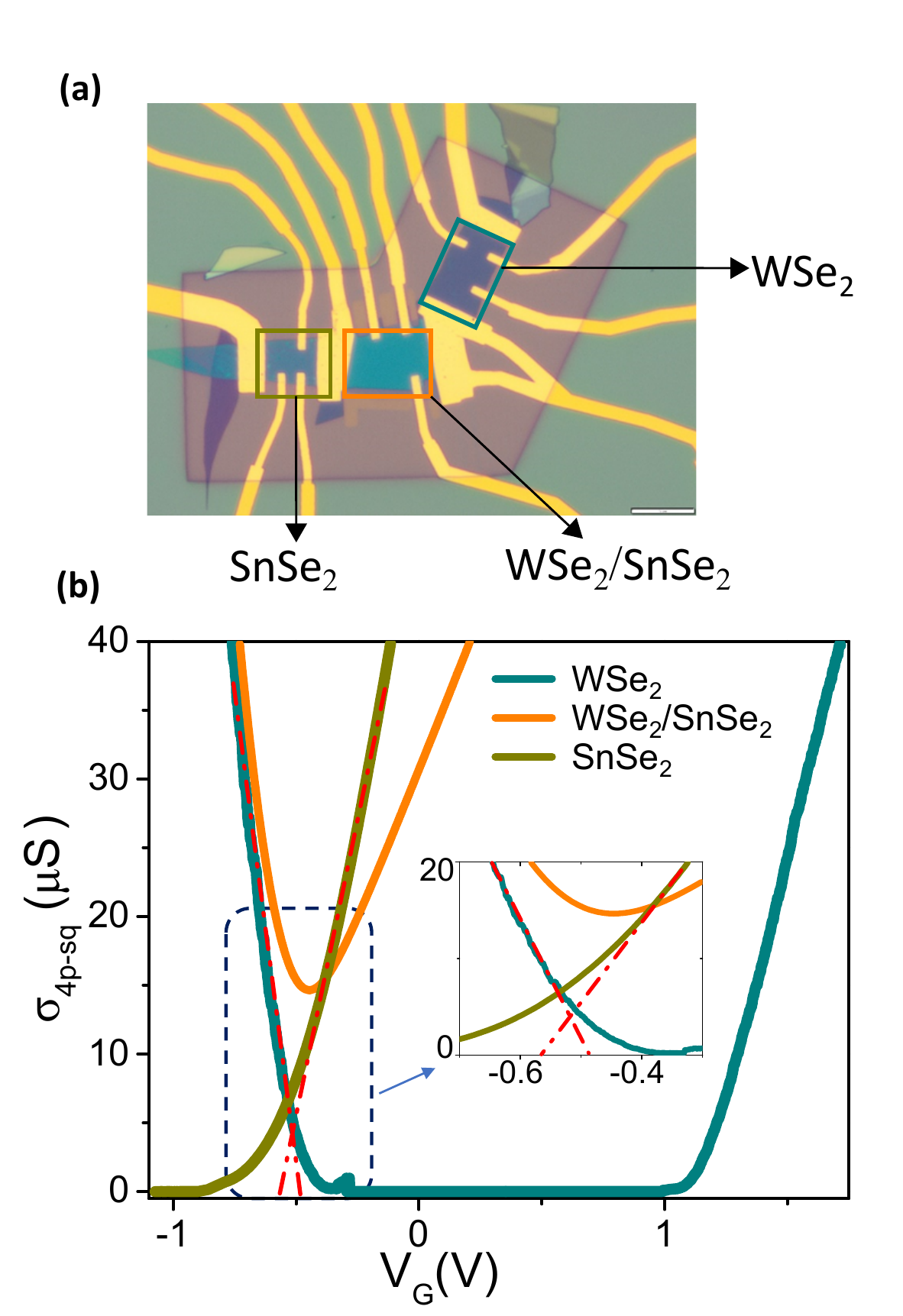}
  \caption{ (a) Optical microscope image of a structure nano-fabricated to enable multi-terminal ionic liquid gate transistor measurements to be performed separately on the interface and on the two constituent materials. The different parts of the structures are indicated by the arrows. The WSe$_2$ and the SnSe$_2$ crystals are respectively three and four layers thick; the SnSe$_2$ layer was capped using a monolayer h-BN to avoid material degradation (the scale bar is 10 $\mu$m). (b) Four-terminal conductivity ($\sigma_{4p}$) measured as a function of ionic-liquid gate voltage ($V_G$) on the individual WSe$_2$ (cyan curve) and SnSe$_2$ (olive curve) layers, as well as on their interface (orange curve). Also in this case, $\sigma_{4p}(V_G)$ shows the characteristic, gap-less behavior of a semimetal (a minimum of conductivity, with the conductivity remaining finite for all $V_G$ values). The red dash-dotted lines indicate how the $\sigma_{4p}(V_G)$ curves are  extrapolated to zero conductivity to extract the relevant threshold voltages. It can be seen that the threshold voltage for hole conduction in WSe$_2$ is larger than the threshold voltage for electron conduction in SnSe$_2$, providing a direct observation of the presence of an interfacial band overlap. Measurements were done at room temperature to avoid freezing of the ionic liquid.}
  \label{fig:Fig4}
\end{figure}

With the aim to probe the electronic properties of WSe$_2$/SnSe$_2$ interfaces more in detail, we fabricated a structure enabling the two constituent  layers and their interface to be probed separately with multiterminal measurements done on a same device. The structure consists of two large exfoliated crystals of 3L-WSe$_2$ and a 4L-SnSe$_2$ that are partially stacked on top of each other, in such a way that also the two constituent layers can be contacted (see Fig. 4(a)). For both layers and for the interface the nano-fabricated contacts do not only enable the longitudinal conductivity to be measured, but also the Hall resistance. All parts of the structure are immersed into a droplet of ionic liquid that covers a common gate electrode. To avoid the possibility of material degradation, we took care to cover with a hBN monolayer the part of the SnSe$_2$ layer not covered by  WSe$_2$.

FET transfer curves measured on the two individual layers and on their interface are plotted in Fig.4b. The transfer curve of  WSe$_2$ (cyan line) exhibits a full gap, as already discussed for  the monolayer case (Fig. 2(e); being thicker than a monolayer, the gap in 3L-WSe$_2$ is smaller, as expected). For the SnSe$_2$ part of the device, the transfer curve (olive line) exhibits conventional behavior, with a clear onset for electron transport below which the conductivity vanishes  (not being relevant here, we did not apply a negative gate voltage large enough to accumulate holes, to avoid risking device degradation). In line with the behavior shown in Fig. 2(e) and Fig. 3, the interface transfer curve (orange line) exhibits a minimum, with the conductivity remaining finite for all value of $V_G$. For $V_G \leq -0.5$V, the $\sigma_{4p}(V_G)$ curve measured on the interface overlaps almost perfectly  with the corresponding curve measured on WSe$_2$, showing that in this range of voltages, the interfacial conductivity originates from holes accumulated in WSe$_2$.  For $V_G > -0.5$V the interfacial conductivity increases upon increasing $V_G$, and its value remains close to that measured on the SnSe$_2$ part of the device, consistently with the notion that in this gate voltage range interfacial transport is mediated by electrons accumulated in SnSe$_2$. The absolute value of the interface conductivity, however, is slightly smaller than in the individual SnSe$_2$ layer, and so is the slope with which $\sigma_{4p}$ increases for larger positive $V_G$. The difference is likely due to the capacitance between the liquid and the SnSe$_2$ layer, which is smaller in the interface region, since there the SnSe$_2$ layer is separated from the liquid by a 3L-WSe$_2$, considerably thicker than the monolayer hBN used to protect SnSe$_2$ in the region away from the interface. The data therefore indicate that transport in the interface region is mediated by holes in WSe$_2$ for $V_G$ values sufficiently smaller than $-0.5$V and by electrons in SnSe$_2$ for $V_G$ values sufficiently larger than $-0.5$V, with the conductivity remaining finite for all $V_G$ values in between (i.e., without vanishing over an extended interval as it happens in the presence of a band gap).

\begin{figure}
    \centering
    \includegraphics[width= 0.6\textwidth]{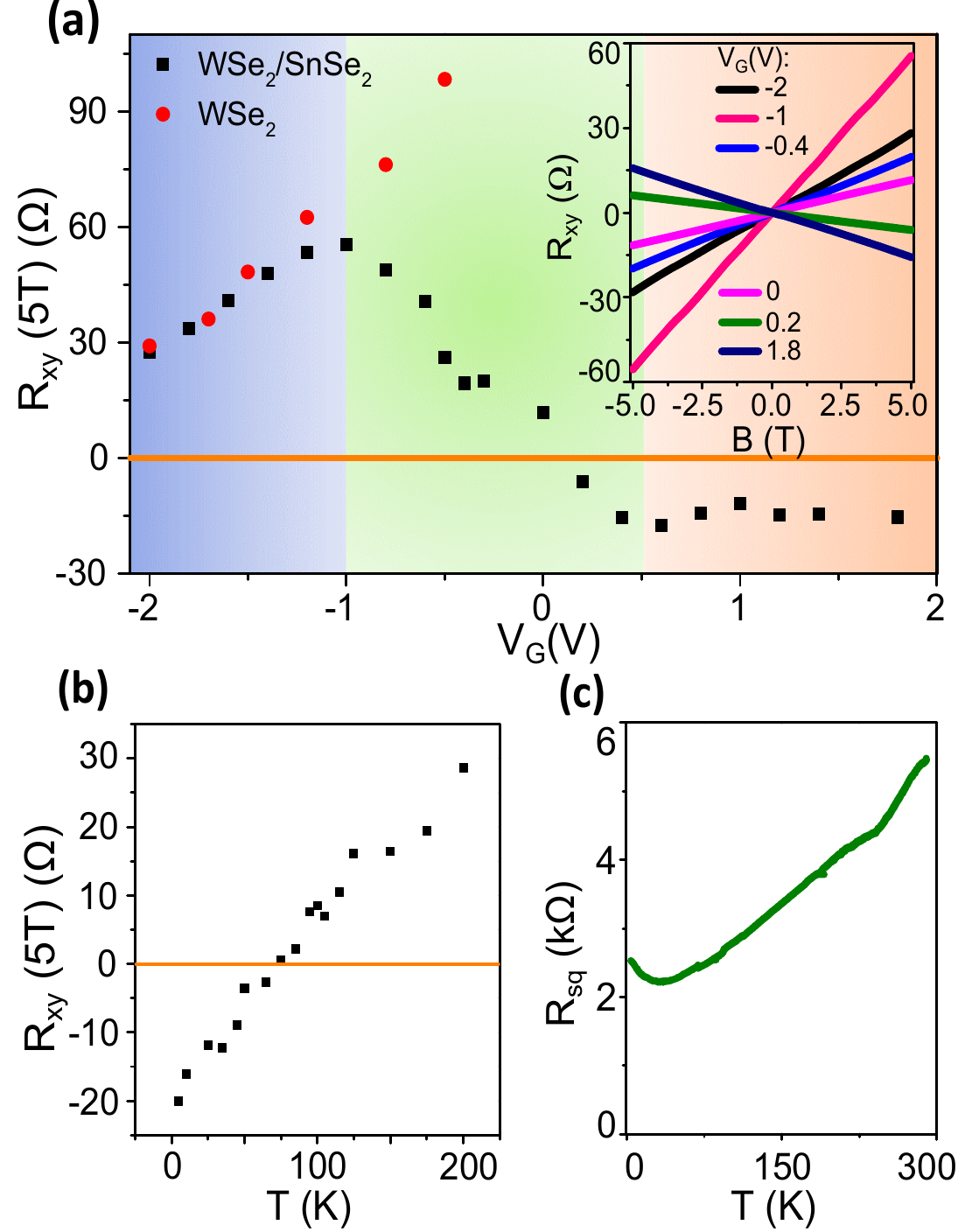}
  \caption{Indications of semimetalliciy in WSe$_2$/SnSe$_2$ interfaces are found in Hall resistance measurements performed at room temperature, as well as in the temperature dependence of the longitudinal resistance. (a) $V_G$ dependence of the transverse resistance $R_{xy}$ measured at $B = 5$T on a 3L-WSe$_2$/4L-SnSe$_2$ interface (black squares; selected traces showing the Hall resistance as a function of  $B$ are shown in the inset) and on the WSe$_2$ layer (red circles; measurements done at room temperature). For WSe$_2$ $R_{xy}$ is positive and increases monotonically with increasing $V_G$, as expected for hole transport. In the interface, $R_{xy}$ initially matches the data measured on the WSe$_2$ layer, but it exhibits a non-monotonic behavior at larger $V_G$ values, eventually reversing sign with further increasing $V_G$. The overall evolution of $R_{xy}$ identifies three regimes --as expected for a semimetal-- shaded with different colors in (a). For $-2$V to $-1$V (blue shaded region) transport is dominated by holes, for $-1$V to $0.5$V (green shaded region) electrons and holes coexist, and for $-0.5$V to $-2$V (orange shaded region) electrons dominate transport. (b) Temperature dependence of $R_{xy}$ in a 2L-WSe$_2$/6L-SnSe$_2$ interface measured at $B = 5$T and fixed $V_G$, in the regime where holes and electrons coexist. $R_{xy}$ changes sign from positive to negative at $T\simeq 85$K, indicating that as $T$ is lowered the dominating type of charge carrier changes from holes to electrons. (c) Temperature dependence of the square resistance $R_{sq}$ of a 2L-WSe$_2$/6L-SnSe$_2$ interface, measured at fixed $V_G$ corresponding to the minimum of conductance. $R_{sq}$ decreases with lowering temperature down to 4K, exhibiting a metallic temperature dependence.}
  \label{fig:Fig5}
\end{figure}

At a more quantitative level, in this device we can determine the threshold voltage $V_{th-SnSe2}^{e}$ for electron conduction in the SnSe$_2$ and the threshold voltage $V_{th-WSe2}^{h}$ for hole conduction in the WSe$_2$, by extrapolating to zero the $V_G$-dependent conductivity of the respective individual layers (see red dashed-dotted lines in Fig. 4(b)). As explained above, the threshold voltages correspond to having the chemical potential located respectively at the conduction band edge of SnSe$_2$ or at the valence band edge of WSe$_2$. We see in Fig. 4(b) that $V_{th-WSe2}^{h}> V_{th-SnSe2}^{e}$, allowing us to conclude directly from the experiments that the valence band edge of WSe$_2$ is higher in energy than the conduction band edge of SnSe$_2$, i.e., that the band alignment of the two materials does exhibit a band overlap. The data also provide an estimate of the magnitude of this overlap, given by $e(V_{th-WSe2}^{h}- V_{th-SnSe2}^{e}) \simeq 80$meV (with a relatively large uncertainty, originating from the extrapolation procedure).

Having concluded that 3L-WSe$_2$/4L-SnSe$_2$ interfaces are indeed  semimetals, we search for the expected manifestations of the semimetallic state in their magnetic-field ($B$) and temperature ($T$) dependent transport properties. We first look at the Hall resistance measured as a function of gate voltage (see black squares in Fig. 5(a)). Starting from large negative $V_G$ values, i.e., when the Fermi level is in the valence band of the interface, the Hall resistance measured at fixed applied magnetic field $B = 5$T (extracted from the curves shown in the inset of Fig. 5(a)) is positive, and initially increases upon shifting $V_G$ to less negative values (corresponding to a decrease in hole density). In this $V_G$ interval the Hall resistance coincides with the one measured on the WSe$_2$ layer (red circles in Fig. 5(a)), confirming that transport in the interface is mediated by states in the valence band of WSe$_2$. Shifting  $V_G$ past $-1$V, causes the Hall resistance to decrease, because in this interval of gate voltage both holes and electrons are present. Indeed, upon increasing $V_G$ past approximately $0$V, the Hall resistance changes sign and becomes negative. Increasing $V_G$ further on the positive side causes the Hall resistance to decrease only very slowly, likely because of the lower capacitance of the ionic liquid to the SnSe$_2$ layer (as we already mentioned, the ionic liquid is separated from SnSe$_2$ by the 3L-WSe$_2$ that is relatively thick) and because exfoliated SnSe$_2$ layers are unintentionally electron doped (i.e., electrons are naturally present in the material at room temperature even in the absence of any applied gate voltage, see Fig. 4(b)). Irrespective of these details, the overall evolution of the Hall resistance as a function of $V_G$ shows the coexistence in the interface of electrons and holes in an extended range of gate voltages \bibnote{To avoid any possible ambiguity we remark explicitly that --contrary to the case when the Fermi level is inside the gap of a semiconducting material-- if the chemical potential lies within the band overlap region  $e\delta V_G \neq \delta E_F$  (because the term $\delta \phi = \frac{e^2 \delta n}{C_G}$ in Eq.(1) cannot be neglected). That is why the $V_G$ interval over which electron and holes coexist (green shaded region in Fig. 5 (a)) is much larger than 80 meV, i.e., the value of band overlap estimated earlier}, which is precisely the behavior expected for a semimetal. The presence of both electrons and holes further manifests itself  in the temperature evolution of the Hall resistance, as shown in Fig. 5(b) for the 2L-WSe$_2$/6L-SnSe$_2$ interface whose transfer curve is shown in Fig. 3. Upon lowering the temperature at $V_G = -1$V, the Hall resistance changes from positive to negative  upon cooling, as the dominating contribution to transport changes from being given by holes at high temperature and by electrons at low temperature. This observation confirms that electrons and holes indeed coexist at the interface.

The presence of a semimetallic state is confirmed by looking at the temperature dependence of the longitudinal square resistance ($R_{sq}$) of the WSe$_2$/SnSe$_2$ interfaces. Fig. 5(c) shows the square resistance measured on a 2L-WSe$_2$/6L-SnSe$_2$ interface gate-biased at the conductance minimum. The square resistance decreases from just over 5k$\Omega$ at room temperature to approximately 2.5k$\Omega$ at $T = 4.2$K, i.e. interfaces gate-biased at their minimum conductivity exhibit a metallic temperature dependence of their resistivity as expected for a semimetal (despite the fact that the interface is formed by two semiconducting materials). We observed metallic temperature dependence in all our devices on which multi-terminal measurements were performed, including devices in which the $T$-dependent resistivity measurements were performed before placing the ionic-liquid, i.e., on the as-assembled structures. For some structures --including 1L-WSe$_2$/1L-SnSe$_2$ interfaces-- we only realized two-terminal devices, not suitable to perform reliable $T$-dependent resistivity measurements, because  of the influence of the contact resistance (which can be large depending on the sample). Even if for these interfaces we cannot entirely exclude that a small gap (rather than a band overlap) is present, all measurements from which a firm conclusion can be drawn (i.e., the gate voltage dependence of both the longitudinal and transverse resistance and the temperature dependence of the longitudinal resistance in all multi-terminal devices investigated) systematically indicate the occurrence of a semimetallic state at WSe$_2$/SnSe$_2$ interfaces.

We emphasize that all the transport properties that we observed in WSe$_2$/SnSe$_2$ interfaces  --the presence of a finite minimum of conductance, the change of sign of the Hall resistance as a function of both gate voltage and temperature, and the metallic temperature dependence of the longitudinal resistance at the minimum of conductance-- are distinctly different from those of the two constituent layers, which are large band gap semiconductors. In the absence of any type of charge accumulation (i.e., due to chemical doping or to a suitably applied gate voltage) WSe$_2$ is an insulator. Even when gate-doped at sufficiently high carrier density to exhibit metallic transport properties, WSe$_2$ never exhibits any signature of spatially coexisting electrons and holes. Similar considerations hold true for SnSe$_2$. Past experiments\cite{zeng_gate-induced_2018} --as well as our work-- find that exfoliated SnSe$_2$ layers are unintentionally doped with a rather high density of electrons, and hence show a rather high room-temperature conductivity at $V_G$= 0V. However, at sufficiently low temperature the square resistance exhibits an exponential increase (shown in Fig. S2(a)). Additionally, as shown in Fig. 4b, gating allows atomically thin SnSe$_2$ layers to be completely depleted, illustrating clearly the presence of a gap. This comparison between the WSe$_2$/SnSe$_2$ interfaces and the individual WSe$_2$ and SnSe$_2$ layers  perfectly illustrate the concept of "synthetic semimetal": WSe$_2$/SnSe$_2$ vdW interfaces truly are  a new system behaving in all regards as a semimetal, despite being formed by two atomically thin crystals that are themselves large-gap semiconductors.

In summary, we have realized interfaces based on atomically thin layers of WSe$_2$ and SnSe$_2$ and shown that they are semimetals. As semimetallic systems can host a variety of interesting physical phenomena depending on microscopic details of their electronic structure, the possibility to study them using vdW interfaces is particularly attractive. Indeed, interfaces offer an experimental control that is not available in naturally occurring semimetals. For instance, the spatial separation of electron and holes --hosted in the two different layers forming the interface-- enables the system to be tuned continuously through the application of a perpendicular electric field that shifts the relative electrostatic potential of the two layers, and to control the overlap between valence and conduction band (to either increase it or suppress it). That is why synthetic semimetals based on van der Waals interfaces of the type demonstrated in our work represent an interesting platform with very considerable potential to reveal new physical phenomena.

\begin{acknowledgement}
We gratefully acknowledge A. Ferreira for continuous technical support and H. Henck for his assistance with the preparation of the figures. A.F.M. gratefully acknowledge financial support from the Swiss National Science Foundation and from the EU Graphene Flagship project.
\end{acknowledgement}

%


\begin{mcitethebibliography}{52}
\bibitem[Not()]{Note-1}
To avoid any possible ambiguity we remark explicitly that --contrary to the
  case when the Fermi level is inside the gap of a semiconducting material-- if
  the chemical potential lies within the band overlap region $e\delta V_G \neq
  \delta E_F$ (because the term $\delta \phi = \frac{e^2 \delta n}{C_G}$ in
  Eq.(1) cannot be neglected). That is why the $V_G$ interval over which
  electron and holes coexist (green shaded region in Fig. 5 (a)) is much larger
  than 80 meV, i.e., the value of band overlap estimated earlier\relax
\mciteBstWouldAddEndPuncttrue
\mciteSetBstMidEndSepPunct{\mcitedefaultmidpunct}
{\mcitedefaultendpunct}{\mcitedefaultseppunct}\relax
\EndOfBibitem
\end{mcitethebibliography}
\providecommand{\latin}[1]{#1}
\makeatletter
\providecommand{\doi}
  {\begingroup\let\do\@makeother\dospecials
  \catcode`\{=1 \catcode`\}=2 \doi@aux}
\providecommand{\doi@aux}[1]{\endgroup\texttt{#1}}
\makeatother
\providecommand*\mcitethebibliography{\thebibliography}
\csname @ifundefined\endcsname{endmcitethebibliography}
  {\let\endmcitethebibliography\endthebibliography}{}

\end{document}